\documentclass[prl,twocolumn,a4paper,amsmath,floatfix]{revtex4}
\usepackage[utf8]{inputenc}
\usepackage{graphicx}
\usepackage{amsmath}

\usepackage{color}
\newcommand{\HexL}{Hex\textsubscript{L}}

\begin{document}
\title{A hard-sphere quasicrystal stabilized by  configurational entropy}
\author{Etienne Fayen$^1$, Laura Filion$^2$, Giuseppe Foffi$^1$, and Frank Smallenburg$^1$}
\affiliation{$^1$Universit\'e Paris-Saclay, CNRS, Laboratoire de Physique des Solides, 91405 Orsay, France\\
$^2$Soft Condensed Matter, Debye Institute of Nanomaterials Science, Utrecht University, Utrecht, Netherlands}

\begin{abstract}
Due to their aperiodic nature, quasicrystals are one of the least understood phases in statistical physics. One significant complication they present in comparison to their periodic counterparts is the fact that any quasicrystal can be realized as an exponentially large number of different tilings, resulting in a significant contribution to the quasicrystal entropy. Here, we use free-energy calculations to demonstrate that it is this configurational entropy which stabilizes a dodecagonal quasicrystal in a binary mixture of hard spheres on a plane.  Our calculations also allow us to quantitatively confirm that in this system all tiling realizations are essentially equally likely, with free-energy differences less than 0.0001$k_BT$ per particle -- an observation that could be the related to the observation of only random tilings in soft matter quasicrystals. Owing to the simplicity of the model and its available counterparts in colloidal experiments, we believe that this system is a excellent candidate to achieve the long-awaited quasicrystal self-assembly on the micron scale.
\end{abstract}
\maketitle

Hard sphere have played a foundational role in our quest to understand classical phase behavior -- from helping to understand how purely entropic systems can crystallize, to revealing new insights into the behavior of glassy materials, to nucleation, to melting in 2d, and many more \cite{royall2023colloidal}.  Their success as a model system stems partly from their inherent simplicity, making them amenable to efficient simulations and analytical theories. Moreover, advances in colloidal particle synthesis have largely made it possible to quantitatively test theoretical and numerical predictions in the lab.

Until recently, quasicrystals were one of the few states of matter inaccessible by this simple model system.  Quasicrystals are exotic structures which can display symmetries that are forbidden to periodic crystal phases.  While highly controversial when first discovered, their place in material science is now well established, with their formation demonstrated in a growing number of both atomic \cite{shechtman1984, tsai1987, tsai2013, bindi2009, bindi2021} and colloidal \cite{dotera2011, zeng2004, hayashida2007, talapin2009, gillard2016, fischer2011a, xiao2012} systems. Toy models that display quasicrystalline behavior have generally been fairly complex -- many early models made use of non-additive binary mixtures of Lennard-Jones particles \cite{widom1987, leung1989} and oscillatory interaction potentials \cite{dzugutov1993formation, dmitrienko1995oscillating, kiselev2012, engel2007self, engel2015}, while more recent work has also explored patchy particles \cite{vanderlinden2012, gemeinhardt2019, noya2021, tracey2021}, anisotropic interactions \cite{haji-akbari2011,haji2009disordered}, and step-wise interactions \cite{dotera2014, pattabhiraman2015}.  Recently, however, we demonstrated the spontaneous self-assembly of two quasicrystal structures in binary mixtures of hard spheres on a plane~\cite{fayen2020, fayen2023}.  This work opens the door to exploring the statistical physics of quasicrystals without the added complication of energetic interactions or orientational degrees of freedom -- in a system that should be realizable in colloidal experiments\cite{thorneywork2014, thorneywork2017}.  One major question in the study of quasicrystals is the role of configurational entropy in their stability\cite{lifshitz2007, barkan2011}. When systems such as hard spheres form quasicrystals, does this happen because the quasicrystal structure maximizes the freedom of particles to vibrate around their quasicrystalline lattice position?  Or are they stabilized by the configurational entropy associated with the large number of possible quasicrystal realizations?

Here, using computer simulations and free-energy calculations, we show that the dodecagonal quasicrystal formed by hard spheres on a plane is stabilized by configurational entropy.  In fact, without the configurational entropy the quasicrystal would be metastable with respect to a phase separation of periodic crystals.  Instead the configurational entropy promotes a random tiling quasicrystal where  -- for this simple hard sphere model -- all realizations contribute equally to the free energy. 


\begin{figure}
    \centering
    \includegraphics[width = \linewidth]{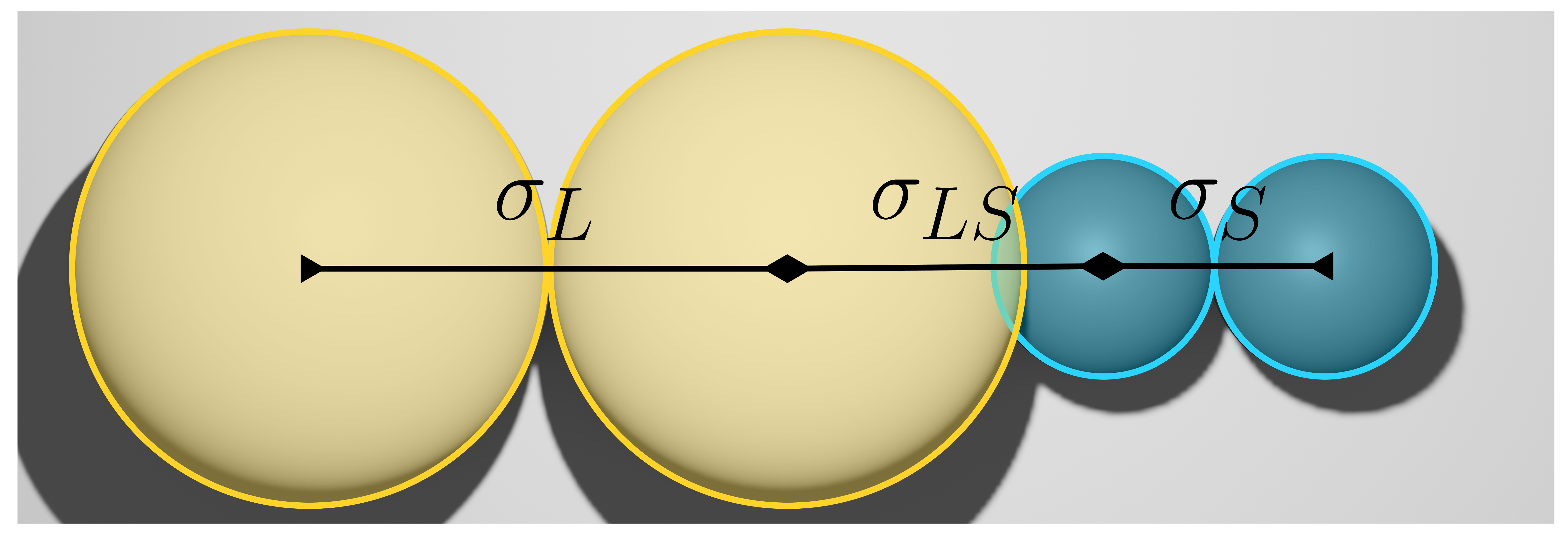}
    \caption{Schematic image of binary hard spheres lying on a flat plane. The line segments indicate the contact distances for the different species.}
    \label{fig:model}
\end{figure}

As illustrated in Fig. \ref{fig:model} we consider binary mixtures of hard spheres constrained to lie on a flat substrate. We focus on systems with a size ratio $q = \sigma_S / \sigma_L$, where $\sigma_{S(L)}$ denotes the diameter of the small (large) spheres. Due to the confinement to a flat plane, the particles can move only in two dimensions, and hence in practice we simulate an effective mixture of non-additive hard disks, where the minimum distance of approach between two disks of unequal size is given by $\sigma_{LS} = \sqrt{\sigma_S \sigma_L}$. Such mixtures are characterized by the composition $x_S = N_S / N$, with $N_{S(L)}$ the number of small (large) spheres and $N$ the total number of spheres. The last free parameter in this model is the packing fraction, which we define as $\eta = (N_S \sigma_S^2 + N_L \sigma_L^2)\pi / 4A$, with $A$ the (two-dimensional) volume of the system. Note that since our binary hard-disk mixture is non-additive the total packing fraction might exceed 1 in some cases. 

Previous work showed that a dodecagonal quasicrystalline phase (QC12) is stable at infinite pressure in this system  for size ratios in the range $0.46 \lesssim q \leq 0.5$\cite{fayen2020, fayen2023}. Moreover, the quasicrystal also forms spontaneously in self-assembly simulations, demonstrating that it is kinetically accessible at finite pressures \cite{fayen2023}. However, this does not prove the thermodynamic stability of this phase, as it could still be metastable with respect to competing periodic crystal phases.  Here, we perform free-energy calculations to settle this question. We focus on mixtures with a size ratio $q = 0.46$. 
Since the QC12 phase only appears for compositions $x_S < 0.5$, we only consider systems with compositions $x_S \leq 0.5$. Interestingly, binary mixtures of hard spheres, not constrained to a plane, have been explored by DFT for size ratios around $\sim 0.8$ in a search for quasicrystals with icosahedral symmetry, which turned out not to be stable \cite{cataldo1999}.

\begin{figure}
	\centering
	\includegraphics[width=\linewidth]{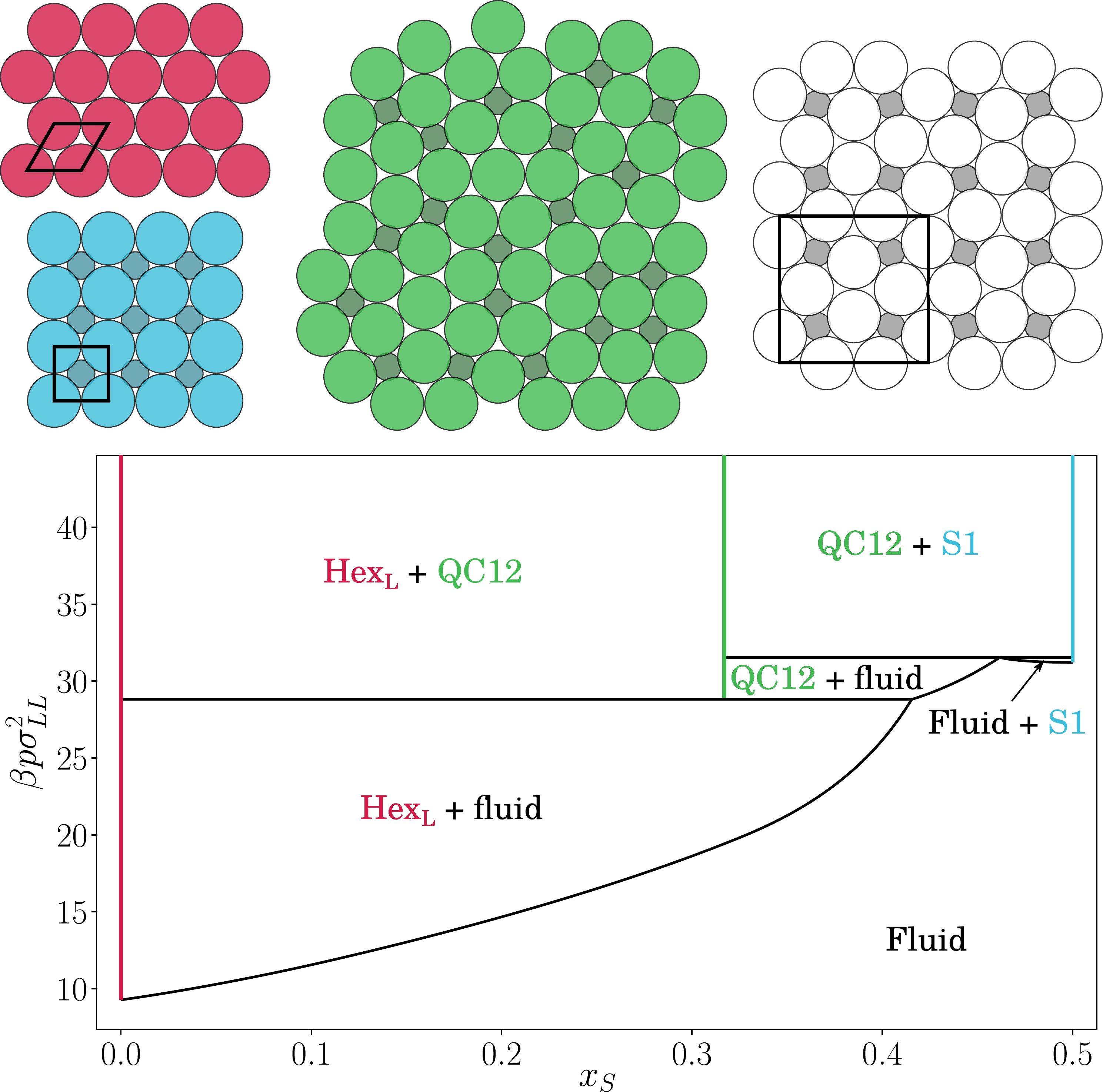}
	\caption{
        (Top) Candidate phases used in the phase diagram construction: hexagonal packing (red), S1 (blue), QC12 (green) and Sigma (white). The unit cell of the periodic structures are depicted as black rhombi.
		(Bottom) Phase diagram of binary mixtures of non-additive hard disks with size ratio $q = 0.46$ corresponding to the equivalent 3D geometry of spheres sedimented on a flat surface. The random tiling dodecagonal quasicrystal is labelled ``QC12''. Although considered as a candidate phase, the Sigma approximant of the quasicrystal is nowhere stable.
	}
	\label{fig:finite_pressure_pd}
\end{figure}

To prove the thermodynamic stability of the QC12 phase, we use explicit free-energy calculations based on both event-driven molecular dynamics simulations \cite{smallenburg2022} and Monte Carlo simulations \cite{frenkel2002a}. In particular, we calculate the free energy of different competing phases as a function of the pressure and composition using thermodynamic integration methods \cite{frenkel2002a}. For the fluid phase, we use the ideal gas as a reference state. For the periodic crystal phases, we obtain a reference free energy using the Einstein molecule variant \cite{vega2007, vega2008} of the Frenkel-Ladd method \cite{frenkel1984}. As candidate structures, we consider the phases that are expected to be stable (or nearly stable) at infinite pressure, namely the hexagonal, S1, Sigma, and QC12 phases \cite{fayen2020}. The candidate phases are depicted in Figure \ref{fig:finite_pressure_pd}.

Determining the stability of a quasicrystal using computer simulations presents challenges that are not present for other crystal phases. First, quasicrystals are non-periodic, and hence the finite-size effects of approximating its aperiodic structure with a periodic approximant should be carefully checked. More importantly, the quasicrystals we expect in colloidal systems are typically examples of so-called \textit{random-tiling quasicrystals}, which results in a configuration entropy contribution to the total free energy of the phase. The quasicrystals of interest here, as well as dodecagonal quasicrystals discovered in soft matter experiments \cite{zeng2004, hayashida2007, talapin2009, gillard2016} and simulations \cite{dotera2011, dotera2014, pattabhiraman2015, widom1987, fayen2023, malescio2022}, are based on a random tiling of the plane by squares and equilateral triangles, with the large particles in the system forming the corners of both shapes. As the number of possible arrangement of these tiles scales exponentially in the number of particles, the freedom of choice in generating this configuration contributes to the total entropy of the phase, and hence needs to be taken into account in any free-energy calculations. This issue is most easily handled if we make the assumption that all realizations of the quasicrystal are equally likely, also known as the random tiling hypothesis \cite{henley1991}. If this is the case, then consistent with the approach of Ref. \onlinecite{pattabhiraman2015} we can split the total free energy of our hard-sphere quasicrystal into two parts:
\begin{equation}
    F_\mathrm{tot}(N,A,T) = F_\mathrm{vib}(N,A,T) - T S_\mathrm{conf},
    \label{eq:RT_hyp_FE}
\end{equation}
where $F_\mathrm{vib}$ is the vibrational free energy of any given quasicrystal realization, $S_\mathrm{conf}$ is the configurational entropy associated with the quasicrystal tiling, and $T$ is the temperature. The vibrational free energy can again be directly calculated for any given realization using the same Einstein molecule approach as we use for the periodic phases.

The configurational entropy of the QC12 square-triangle tiling is well studied \cite{oxborrow1993}. When the ratio of the number of squares $N_{sq}$ and triangles $N_{tr}$ reaches $N_{sq}/N_{tr}=\sqrt{3}/4$, the random tiling ensemble reaches a maximum entropy, meaning that the number of tilings in the ensemble, or equivalently the number of possible configurations for the squares and triangles, is the highest. At this point, the random tiling ensemble forms a so-called \emph{random-tiling quasicrystal} of 12-fold symmetry \cite{widom1993, kawamura1983, kalugin1994, nienhuis1998, imperor-clerc2021}. 
The configurational entropy of the square-triangle tiling was first estimated with transfer matrix \cite{kawamura1983, kawamura1991} and numerical \cite{oxborrow1993} approaches, before exact analytical expressions were obtained with a Bethe ansatz \cite{widom1993,kalugin1994}. Based on these works, the random tiling configurational entropy per particle is given by
\begin{equation}
    S_\mathrm{conf}/Nk_B = [\ln(108) - 2\sqrt{3}\ln{(2+\sqrt{3})}] (1 - x_S^\text{QC12}) \approx 0.082. \label{eq:Sconf}
\end{equation}
Here, $x_S$ is the composition of the system which corresponds to the ratio of squares and triangles required for a quasicrystal, i.e. $x_S^\text{QC12} = \sqrt{3} / (2 + 2\sqrt{3}) \approx 0.317$. Importantly, $S_\mathrm{conf}$ is sharply peaked at this composition, and $-TS_\mathrm{conf}$ is non-convex on either side of the maximum \cite{kalugin1994}, such that random tilings at any compositions other than $x_S^\text{QC12}$ are strongly entropically disfavored.


To explore the stability of the QC12 phase, we construct the phase diagram as a function of the composition $x_S$ and pressure $p$. To this end, we transform the (Helmholtz) free energies obtained from our thermodynamic integration into Gibbs free energies using the equation of state of the respective phases. The coexistence regions are mapped out using common tangent constructions at constant pressure. 
Note that we do not expect phases with $x_S > 0.5$ to play a role in this phase diagram, as the S1 phase (at $x_S = 0.5$) is the best-packed phase for this system and previous self-assembly studies did not show any self-assembly of higher-composition phases at $x_S < 0.5$.

The resulting phase diagram is shown in Fig. \ref{fig:finite_pressure_pd}, and clearly indicates a broad stable region for the QC12 phase. Additionally, we observe a binary S1 solid phase, a hexagonal solid of large particles, and a binary fluid phase. Note that in addition to these phases, we have confirmed that the Sigma phase, which is the first approximant to the dodecagonal quasicrystal, is not stable.
Although the QC12 phase mostly coexist with other solid phases for most of its stability range, there is a narrow band of pressures where it coexists with a fluid with a larger concentration of small particles. As self-assembly is likely to be easier to achieve from a fluid phase, this suggests that self-assembly of this phase may be easiest by starting from an off-stoichiometric fluid with $x_S > x_S^{QC12}$. This scenario is in line with earlier self-assembly observations \cite{fayen2023}, and has previously been reported for other phases as well \cite{lacour2022tuning}.

\begin{figure}
	\centering
	\includegraphics[width=\linewidth]{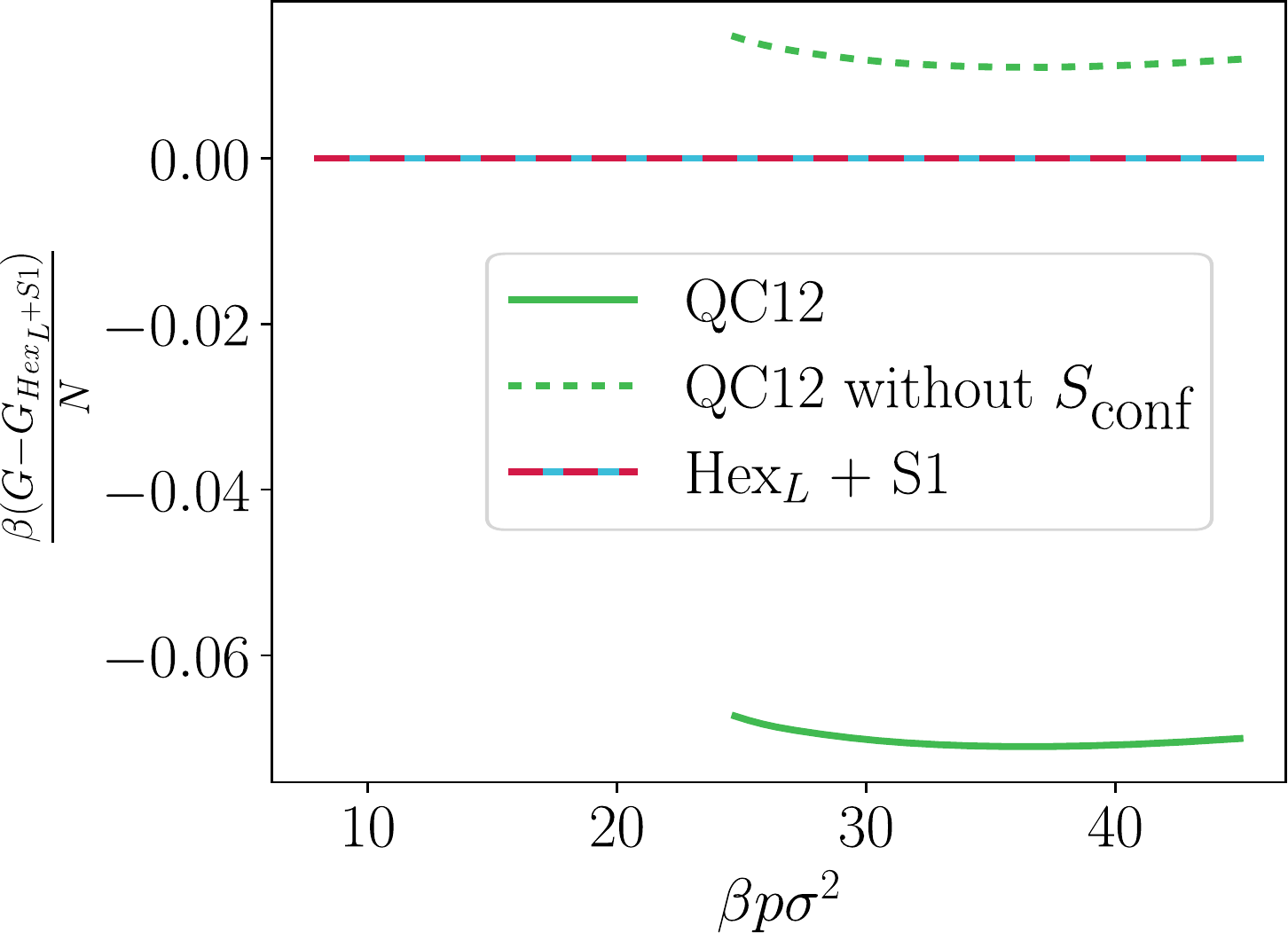}
	\caption{
		Free-energy difference between the competing coexistence of \HexL{} + S1 and the quasicrystal at the quasicrystal composition. For the quasicrystal, the dashed curve corresponding to the vibrational entropy	alone and lies above the coexistence free-energy, while the addition of	the constant tiling entropy term (solid line) stabilizes the quasicrystal.
	}
	\label{fig:qc12_coexistence}
\end{figure}

An interesting question is whether the quasicrystal is stabilized purely by its vibrational entropy, as previously proposed for the same quasicrystal phase in particles interacting via a square-shoulder repulsive potential \cite{pattabhiraman2015}, or whether the configurational entropy is essential for its stability. To check this,  in Figure \ref{fig:qc12_coexistence} we compare the free energies of the dodecagonal quasicrystal and the competing coexistence of the hexagonal and S1 solids at the quasicrystal composition. Without the configurational entropy term, the Hex$_L$-S1 coexistence prevails and the quasicrystal is not stable. Clearly, for this system, the tiling contribution to the total entropy is critical for the quasicrystal stability. Unfortunately, this also implies that we can only conclude that the QC12 phase is stable if we are justified in the assumption that all tiling realizations are equally likely, and hence that Eq. \ref{eq:RT_hyp_FE} is correct. In order to confirm this, we need to check that different configurations of the random tiling ensemble are degenerate in vibrational entropy.

To this end, we perform high-precision calculations of the vibrational entropy of various tiling realizations. In particular, we compare the vibrational entropy of several types of \emph{ideal} quasicrystal configurations, as well as fully randomized tilings. Ideal quasicrystal configurations can be generated by so-called inflation methods, in which every tile of a tiling is replaced by a cluster of tiles. By iterating the inflation rules on an initial seed, one generates larger and larger patches of tiling that converge to a quasicrystalline configuration. Several inflation rules exist for the square-triangle tiling. The most common way of producing a fully deterministic quasicrystal tiling with dodecagonal symmetry is via the Schlottmann inflation rule \cite{hermisson1997, frettloeh2011}. Alternatively, a quasicrystal tiling with hexagonal symmetry can be constructed using the simpler Stampfli inflation rule \cite{stampfli1986}. A slight variation of the Stampfli uses random choices to generate a limited ensemble of random tiling realizations with 12-fold symmetry on average. Finally, configurations from the full random tiling ensemble can be sampled by reshuffling ideal configurations using so-called zipper moves that rearrange tiles along a closed path in the tiling \cite{oxborrow1993}. More details on the generation of our tiling configurations can be found in the Supplemental Information (SI). 

Note that in terms of quasicrystal language, these different types of tilings all have zero perpendicular strain, but differ in terms of the fluctuations of their representative surface in the perpendicular space \cite{henley1991, janot1998, janssen2007}. In particular, the deterministic Schlottmann and Stampfli rules have minimal fluctuations in their representative surface, while the full random tiling ensemble has much stronger fluctuations.

\begin{figure}
	\centering
	\includegraphics[width=\linewidth]{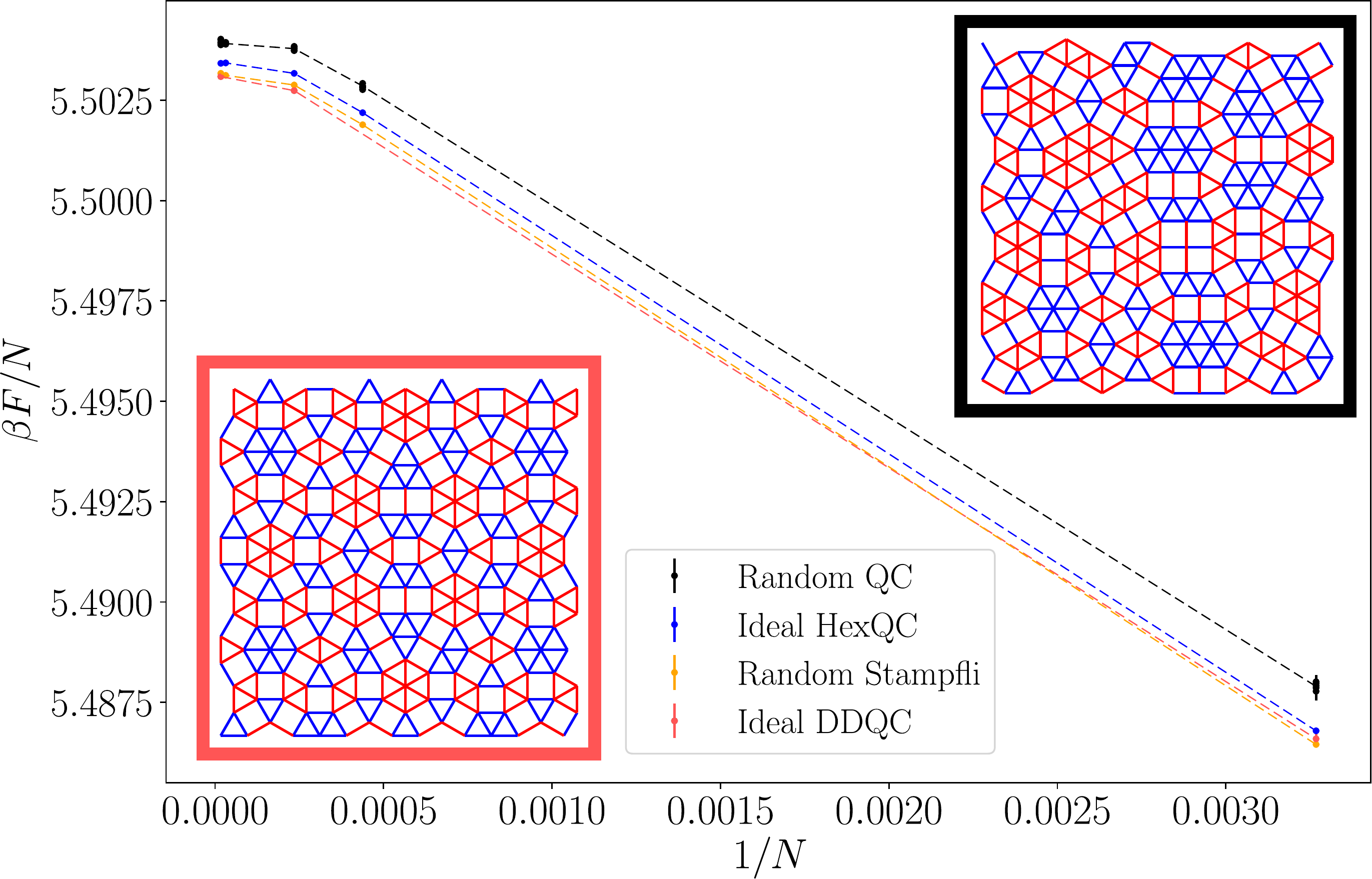}
	\caption{
		Free energies of the Schlottmann (dodecagonal, pink), random Stampfli (dodecagonal, orange), 
		Stampfli (hexagonal, blue) and 5 fully random quasicrystals (dodecagonal, black), for
		various system size. The error bars are the statistical error on the
		mean obtained by repeating Frenkel-Ladd calculations many times, and
		are smaller than the symbols for most points. Two system sizes are
		obtained by inflation of a sigma seed, for which our implementation of
		the Schlottmann inflation fails. Hence, two pink points are missing. The inset show examples of an ideal Schlottmann tiling (bottom left) and a random tiling (upper right).
	}
	\label{fig:ideal_random_QC}
\end{figure}

To test whether the vibrational entropies of these different families of quasicrystal tilings are degenerate, we use again the Einstein molecule approach. We calculate the free energy of configurations from each of these families for several different system sizes. For the randomly generated tilings, we create 5 different random configurations by applying zipper moves to the approximant and calculate the free energy of each. The density is fixed at $1.5\, \sigma_{LL}^{-2}$ for all systems. Note that in order to minimize statistical error and reduce the error bars, we repeat the free-energy calculation for each configuration at least 100 times (see SI) and average over the results.

The results are shown in Fig. \ref{fig:ideal_random_QC}.  The finite size scaling of the free energy appears to be non-linear for each structure, and adding the heuristic finite size correction term $\ln(N)/(2N)$ proposed in Ref. \citenum{frenkel2002a} does not remove the non-linearity. One could argue that a linear regime is reached for very large system sizes, with an almost zero slope. Therefore, we perform no extrapolation and use the value of the free energy per particle for the largest systems as our estimate of the thermodynamic-limit value. We obtain $\beta F/N = 5.50309 (5)$ for the Schlottmann quasicrystal, $5.50317 (4)$ for the random Stampfli quasicrystal, $5.50342 (4)$ for the Stampfli hexagonal quasicrystal and $5.50392 (4)$ for the average over the 5 largest realizations of the random tiling quasicrystal. 

An important first observation is that the free energies of the 5 random-tiling quasicrystals generated at each system size are consistently degenerate within our errorbars (black clusters in Figure \ref{fig:ideal_random_QC}). The absence of any outliers gives us confidence that the vast majority of configurations in the random tiling ensemble indeed have essentially the same vibrational entropy. This observation quantitatively validates the assumption that all realizations are equally likely in our system and justifies the treatment of the QC12 as a random tiling phase with the configurational entropy given by Eq. \ref{eq:Sconf}. 

The measurements show, nonetheless, that some configurations in the random tiling ensemble are special. The free energy of the inflated quasicrystals is consistently lower than that of the random configurations, with the difference on the order of $10^{-3} k_BT$ per particle. This difference is much too small to affect the stability of the random quasicrystal phase, as can be seen by comparing it to the scale of free-energy differences in Fig. \ref{fig:qc12_coexistence}. However, it is measurable, and of the same order of magnitude as the free-energy difference between face-centered cubic (FCC) and hexagonal close-packed (HCP) crystals of monodisperse hard spheres \cite{frenkel1984}. Using a self-consistent field theory, Duan \emph{et. al.} also demonstrated a free-energy difference between ideal and random configurations of a dodecagonal quasicrystal in a system of tetrablock copolymers \cite{duan2018}. Moreover, we find that the ideal dodecagonal quasicrystal obtained with Schlottmann inflation has slightly more vibrational entropy than both the ideal hexagonal Stampfli and random Stampfli quasicrystals, although the difference with the latter is very small. 

The vibrational entropy difference between random and ideal quasicrystals can be understood from the different local environments that can be found in the underlying tiling. For instance, ideal quasicrystals obtained by the inflation method contain no local environment formed of 4 squares meeting at the same vertex while the randomized ones contain a non-zero concentration of those \cite{rubinstein2000} (see Fig. \ref{fig:ideal_random_QC}). We expect however that the first-neighbor local environments alone do not explain fully the entropy difference. Indeed, both the dodecagonal and hexagonal ideal quasicrystals have the same distribution of local environments when considering only the first neighbor shell. Hence, neighbor shells beyond the first one certainly play a non-negligible role. 

From the point of view of quasicrystal theory, the vibrational entropy difference between ideal and random structures is an interesting  illustration of phonon-phason coupling \cite{henley1991, kiselev2012}, albeit very weak. The vibrational entropy of each system can be interpreted as stemming from the total entropy contribution from all phonon modes accessible to the quasicrystal. In this picture, the lower vibrational entropy of the random quasicrystals shows that the presence of phason modes in these systems hinders lattice vibrations, \textit{i.e.} reduces the amplitude of the phonon modes. 


In conclusion, our results demonstrate the thermodynamic stability of a dodecagonal quasicrystal in a binary mixture of hard spheres confined to lie on a flat substrate.  As it consists of hard spheres, the quasicrystal considered here is inherently stabilized by entropy alone. Importantly, however, it is also an example of a quasicrystal that is stabilized by its configurational, rather than vibrational, entropy. This configurational entropy stems from the many different possible tiling realizations, which are -- as shown by our precise free-energy calculations -- nearly indistinguishable in terms of their vibrational freedom.  Due to the tiny free-energy difference between different realizations, random tilings are overwhelmingly more likely to form than perfect inflationary tilings. We speculate that this observation could be related to the fact that, in soft matter, all the quasicrystalline systems observed thus far appear to indeed be random. 
Note, however, that in some systems, sufficiently strong particle interactions could favor or suppress different sets of quasicrystal realizations, lowering the configurational entropy and potentially destabilizing the quasicrystal phase \cite{kiselev2012}.

Additionally, since sedimented systems of hard colloidal spheres can be readily realized in the lab \cite{thorneywork2014, thorneywork2017}, this equilibrium quasicrystal is extremely promising for the creation and study of quasicrystals on the colloidal scale. Such a realization would be an important step forward in the study of (soft-matter) quasicrystals, as it would provide an ideal platform for the real-space study of e.g. defect dynamics, perpendicular strain relaxation, and other phenomena that are hard to study in molecular or atomic quasicrystals.


\section{Acknowledgements}

We thank Anuradha Jagannathan, Marianne Imp\'eror-Clerc, Pavel Kalugin, and Alfons van Blaaderen for interesting and useful discussions. 
EF, GF, and FS acknowledge funding from the  Agence Nationale de la Recherche (ANR), grant ANR-18-CE09-0025.  LF acknowledges funding from the Dutch Research Council (NWO) under the grant number OCENW.GROOT.2019.071.  The authors acknowledge the use of the Ceres high-performance computer cluster at the Laboratoire de Physique des Solides to carry out the research reported in this article.

\bibliography{refs}

\end{document}